\documentstyle[12pt,epsfig]{article}

\begin{document}


\begin{center}
{\bf NEW DATA ON OZI RULE VIOLATION IN $\bar{p}p$ ANNIHILATION AT REST} \\ ~ \\
{\bf The OBELIX collaboration.}\\ 
\end{center}
A.Ber\-tin$^a$, M.Brus\-chi$^a$,
S.De Cas\-tro$^a$, A.Fer\-ret\-ti$^a$,
D.Gal\-li$^a$, B.Gia\-cob\-be$^a$,
U.Mar\-co\-ni$^a$, M.Poli$^{a,}$\footnote
{\it Dipartimento di Energetica, Universit\`a di Firenze},
M.Pic\-ci\-ni\-ni$^a$, N.Sem\-pri\-ni-Ce\-sa\-ri$^a$,
R.Spi\-ghi$^a$, S.Vec\-chi$^a$, A.Vez\-za\-ni$^a$, F.Vi\-got\-ti$^a$,
M.Vil\-la$^a$, A.Vi\-ta\-le$^a$,
A.Zoc\-co\-li$^a$,
G.Bel\-li$^b$, M.Cor\-ra\-di\-ni$^b$, A.D\-on\-zel\-la$^b$,
E.Lo\-di Riz\-zi\-ni$^b$, L.Ven\-tu\-rel\-li$^b$,
A.Ze\-no\-ni$^c$, C.Ci\-ca\-l\'o$^d$, A.Ma\-so\-ni$^d$, G.Pud\-du$^d$,
S.Ser\-ci$^d$,
P.P.Tem\-ni\-kov$^d$, G.U\-sa\-i$^d$,
O.Yu.De\-ni\-sov$^e$, O.E.Gor\-cha\-kov$^e$, V.P.No\-mo\-ko\-nov$^e$,
S.N.Pra\-khov$^e$, A.M.Rozh\-dest\-ven\-sky$^e$, M.G.Sa\-pozhni\-kov$^e$,
V.I.Tre\-tyak$^e$, P.Gia\-not\-ti$^f$, C.Gua\-ral\-do$^f$, A.La\-na\-ro$^f$,
V.Lu\-che\-ri\-ni$^f$, F.Ni\-chi\-tiu$^{f,}$\footnote
{On leave of absence from
{\it Department of High Energy Physics, Institute of Atomic Physics,
Bucharest, Romania}}, C. Pe\-tras\-cu$^{f,2}$,
A. Ros\-ca$^{f,2}$,
V.G.A\-ble\-ev$^{g,}$\footnote {On leave of absence from
{\it Joint Institute for Nuclear Research, Dubna, Russia}},
C.Ca\-vion$^g$, U.Gas\-tal\-di$^g$, M.Lom\-bar\-di$^g$,
A.An\-dri\-ghet\-to$^h$, M.Mo\-ran\-do$^h$,
G.Ben\-di\-scio\-li$^i$, V.Fi\-lip\-pi\-ni$^i$, A.Fon\-ta\-na$^i$,
P.Mon\-ta\-gna$^i$, A.Ro\-ton\-di$^i$, A.Sa\-i\-no$^i$,
P.Sal\-vi\-ni$^i$,
F.Ba\-les\-tra$^j$, E.Bot\-ta$^j$, T.Bres\-sa\-ni$^j$,
M.P.Bus\-sa$^j$,
L.Bus\-so$^j$,
D.Cal\-vo$^j$, P.Ce\-rel\-lo$^j$, S.Cos\-ta$^j$, L.Fa\-va$^j$,
A.Fe\-li\-ciel\-lo$^j$, L.Fer\-re\-ro$^j$,
A.Fi\-lip\-pi$^j$, R.Gar\-fa\-gni\-ni$^j$,
A.Gras\-so$^j$, D.D'I\-sep$^j$, A.Mag\-gio\-ra$^j$, S.Mar\-cel\-lo$^j$,
D.Pan\-zie\-ri$^j$, D.Pa\-re\-na$^j$,
E.Ros\-set\-to$^j$, F.To\-sel\-lo$^j$,
G.Zo\-si$^j$,
M.A\-gnel\-lo$^k$, F.Iaz\-zi$^k$, B.Mi\-net\-ti$^k$,
G.V.Mar\-ga\-gliot\-ti$^l$, G.Pa\-u\-li$^l$, S.Tes\-sa\-ro$^l$,
L.San\-ti$^m$ \\ ~ \\
{$^a$\it Dipartimento di Fisica, Universit\`a di Bologna and
INFN, Sezione di Bologna, Bologna, Italy} \\
{$^b$\it Dipartimento di Chimica e Fisica per i Materiali, Universit\`a di
Brescia and INFN, Sezione di Torino, Torino, Italy} \\
{$^c$\it Dipartimento di Chimica e Fisica per i Materiali, Universit\`a di
Brescia and INFN, Sezione di Pavia, Pavia, Italy} \\
{$^d$\it Dipartimento di Scienze Fisiche, Universit\`a di Cagliari and
INFN, Sezione di Cagliari, Cagliari, Italy} \\
{$^e$\it Joint Institute for Nuclear  Research, Dubna, Russia} \\
{$^f$\it Laboratori Nazionali  di  Frascati  dell'INFN, Frascati, Italy} \\
{$^g$\it Laboratori Nazionali di Legnaro dell'INFN, Legnaro, Italy} \\
{$^h$\it Dipartimento di Fisica, Universit\`a di Padova and INFN, Sezione
di Padova, Padova, Italy} \\
{$^i$\it Dipartimento di Fisica Nucleare e Teorica, Universit\`a di Pavia, and
INFN Sezione di Pavia, Pavia, Italy} \\
{$^j$\it Dipartimento di Fisica, Universit\`a di Torino and
INFN, Sezione di Torino, Torino, Italy} \\
{$^k$\it Politecnico di Torino and INFN, Sezione di Torino, Torino,
Italy} \\
{$^l$\it Istituto di Fisica, Universit\`a di Trieste and  INFN, Sezione  di
Trieste, Trieste, Italy} \\
{$^m$\it Istituto di Fisica, Universit\`a di Udine and
INFN, Sezione di Trieste, Trieste, Italy} \\

\begin{abstract}

  The results of a measurement of the ratio 
$R = Y(\phi\pi^+\pi^-) / Y(\omega\pi^+\pi^-)$ for antiproton
annihilation at rest in a gaseous and in a liquid hydrogen target
are presented.
It was found that the value of this ratio 
increases with the decreasing
of the dipion mass, which demonstrates the difference in the $\phi$ and
$\omega$ production mechanisms. 
An indication on the momentum transfer dependence of the apparent
OZI rule violation for  
$\phi$ production from the $^3S_1$ 
initial state was found.
\end{abstract}

In recent experiments with
stopped antiprotons at LEAR (CERN)
\cite{Rei.91}-\cite{Abl.95b}
strong violation
of the Okubo-Zweig-Iizuka rule
\cite{OZI} was found.
The  ratio of the yields of $\phi$ and $\omega$ mesons $R(\phi X/\omega X)$ 
turned out
about 30- 50 times larger than the OZI prediction
$R(\phi X/\omega X)\approx4\cdot10^{-3}$.
In particular,  the degree
of the OZI rule violation was found \cite{Rei.91},\cite{Abl.95b} 
to depend strongly
on the quantum numbers of the initial state.

  A number of theoretical models were developed
for explaining these data. 
Approaches based on the traditional ideas \cite{Lev.94},\cite{Loc.94} 
are  unable to
reproduce {\it all} the features of $\phi$ production observed now (for
a review, see \cite{Sap.94}).
It has been suggested \cite{Ell.89}-
\cite{Alb.95} that the nucleon wave function contains some $s\bar{s}$
pairs already at small momentum transfer. Then the abundant $\phi$
production in $\bar{N}N$ annihilation  
should not be considered as a violation of the OZI rule,
because it does not involve disconnecting quarks diagram. In this model
the $\phi$ production is described in terms of rearrangment of the
$s\bar{s}$ pairs already stored in the nucleons. 

	It is differ from a standard view that the $\phi$, 
which is practically pure 
$s\bar{s}$ state, 
should be created in the interactions of the non-strange hadrons  
only due to small admixture of the light quarks in
its wave function and the production amplitudes of $\phi$ and $\omega$
mesons should be rather similar.

In this paper, we study the production mechanisms of $\phi$ and
$\omega$ mesons in the  reactions
\begin{eqnarray}
\bar{p} + p & \longrightarrow & \phi + \pi^+ + \pi^- \\
\bar{p} + p & \longrightarrow & \omega + \pi^+ + \pi^-
\end{eqnarray}
of antiproton annihilation at rest. The ratio
$R = Y(\phi\pi^+\pi^-) / Y(\omega\pi^+\pi^-)$
was measured for different invariant masses of the dipion system.
It was found that the value of this ratio  
increases with the 
decreasing of the dipion mass, which demonstrates the difference in the $\phi$
and
$\omega$ production mechanisms. 

\vspace{0.5cm}
   Antiproton annihilations at rest were obtained using
the OBELIX spectrometer on the M2 beam line of LEAR.
A detailed description of the spectrometer can be found elsewhere 
\cite{Ada.92}.
The experimental setup consists of a number of subdetectors arranged
around the Open Axial Field Magnet. The Time Of Flight (TOF) system
contains 
 two coaxial barrels of plastic scintillators 
for charged particle
identification and trigger. The Jet Drift Chamber (JDC) 
provides
geometry reconstruction of the event and particle identification by
energy loss measurement. The High Angular Resolution Gamma Detector
(HARGD)
reconstructs the multiplicity, the direction and the energy of the
detected gammas.

    In order to investigate the reactions (1)-(2),
events
with 4 charged particles were analyzed.
To study 
 $\omega-$meson
production
the trigger, based on the TOF system, was used for selection of events
characterized by 4 hits in the inner barrel of the scintillators and 3-4 hits
is the outer one. 
To study 
 $\phi-$meson
production a sample of data 
was collected using also a "slow particle" condition - at least
one time difference between hits in the inner and outer barrels had to 
be greater than 8 nsec. This condition was used to enrich the sample by
charged kaons.

Data were taken using two hydrogen targets: liquid  and gas
at 3 atm pressure. In the former case, annihilations occur mainly from
S-wave initial states, whilst in the latter one
annihilations from S- and
P-waves are possible with approximately equal probability.
 A sample of 
$16 \cdot 10^6$  
annihilations in liquid target 
 with the "slow particle" condition was collected and 
$1 \cdot 10^6$ events
without it were taken. For the gas target,  $1.5 \cdot 10^6$ and
$3 \cdot 10^6$ events, respectively, were taken.

The following criteria were used for event selection.
The event should have 4 tracks with total charge zero.
Data samples with the "slow particle"
trigger were used to select events with $K^+K^-\pi^+\pi^-$ 
final state. Two kaons of opposite charge should be
recognized by dE/dx and TOF. Finally, the events were submitted to kinematical
fit to fulfill the hypotheses
$ \bar p p \to K^+K^-\pi^+\pi^-$ (4C fit) at $CL=5\%$
and
$ \bar p p \to \pi^+\pi^-\pi^+\pi^-\pi^0$ (1C fit) at 
 $CL=10\%$.

For the liquid target sample, 59299 $KK\pi\pi$ events and 127904 $5\pi$ events
were selected. For the gas target sample,
7015 and 230769 events, respectively, were selected.

\vspace{0.5cm}
     The $K^+K^-$ and $\pi^+\pi^-\pi^0$ invariant mass spectra for the selected
events are shown in Fig.1. Clear signals from $\phi$ and
$\omega$ mesons are apparent.
Approximation of them by the sum of
a Breit-Wigner function with fixed $\phi(\omega)$ width, smeared by a Gaussian
distribution gives 
the following values for mass and experimental resolution $\sigma$
(for the gas target sample): 
$m_{\phi}=(1019.1\pm0.5)~MeV/c^2$, 
$\sigma_{\phi} = (5.3\pm0.6)~MeV/c^2$ and
$m_{\omega}=(781\pm1)~MeV/c^2$, 
$\sigma_{\omega} = (18.9\pm0.1)~ MeV/c^2$.

 Background free $\pi^+\pi^-$ invariant mass distributions
(uncorrected for apparatus acceptance) for the reactions (1) and (2)
are shown in Fig.2 a)-d).
For the reaction (1)  ($\phi$ production)
the spectra were obtained after subtraction
of events from the bins  nearby
the $\phi$ peak.

For the reaction (2) ($\omega$ production),
so-called
$\lambda$-subtraction procedure\cite{Biz.69} was used.
The $\omega$ decay amplitude is proportional to the
parameter
$$\lambda=
\frac{|\vec p_{\pi^-}\times \vec p_{\pi^+}|^2}{\lambda '}$$
 where
$\vec p$ are the three-momenta of the charged pions  from
the $\omega$ decay in its rest frame, and $\lambda'$
is the maximum possible value of $|\vec p_{\pi^-}\times \vec p_{\pi^+}|^2$.
The distribution of this parameter for events coming from  $\omega$ decay is
proportional to $\lambda$, while for the background events 
the corresponding $\lambda$--distribution should
be flat. This was verified for the analyzed sample. 
The events were separated into two samples: one with $\lambda<0.5$
and the other with $\lambda>0.5$. Subtracting the spectrum with
$\lambda<0.5$ from that $\lambda>0.5$ one can obtain a background-free
distribution. 

In the $\pi^+\pi^-$  invariant mass distributions associated to $\phi$
production
shown in Fig.2 a,b),  there is no evidence of a  
$\rho$-meson peak. This is consequence
of the apparatus acceptance. Charged kaons with a momentum less than 200 MeV/c
cannot arrive at the JDC. As a result, it is not possible to observe high
values of $M_{\pi\pi}$ created together with $\phi$.

The dipion mass distributions associated to $\omega$ production
are shown in Fig.2 c,d).
They are dominated by the peak
of $\rho$ meson production, which turns out especially evident for annihilation
in gas. In the case of liquid target (Fig. 2c), 
remnants of the $f_2(1270)$ peak, truncated by the phase space, can be seen.
These distributions are rather similar to those measured in the previous
experiments of $\omega$ production  for annihilation in liquid
\cite{Biz.69} and gas \cite{Wei.93} targets.
From  Figs.2 c,d) it turns out that at small
masses, $300~MeV/c^2<M_{\pi\pi} < 500~MeV/c^2$,  
the invariant mass distributions
are rather flat. This interval is also
suitable for detection of
pions created together with $\phi$ mesons in the reaction (1).

The ratio
$R = Y(\phi\pi\pi)/Y(\omega\pi\pi)$ (where Y is the yield of  $\phi$- or
$\omega$-mesons) was determined 
using the numbers of $\phi$ and
$\omega$  obtained from the $K^+K^-$ and $\pi^+\pi^-\pi^0$ invariant
mass distributions
for dipion  mass within the interval
$300~ MeV/c^2 < M_{\pi\pi} < 500~ MeV/c^2$ and compared  with the same
ratio obtained for all events.   
To avoid a loss of useful events, 
we not used the background-free distributions. 
To estimate the number of $\phi$($\omega$)
mesons the corresponding part of the invariant mass spectrum
was approximated by the sum of
a Breit-Wigner function with fixed $\phi(\omega)$ width, smeared by a Gaussian
distribution and a
polynomial function to describe the background.

To evaluate the detection efficiency 
of reactions (1)-(2),  
Monte-Carlo simulated
events were reconstructed and analysed using the same selection criteria
and kinematical cuts as for real data.
The phase space distribution of the dipion mass was assumed
both in case of $\phi$ and $\omega$
production. 

In Table, the number of events
$N_{ev}$ selected for each reaction, the detection efficiencies $\epsilon$,
the yields $Y$ of $\phi$ and $\omega$ meson production  and the
corresponding ratios R are reported. 
The yield is calculated as
$Y = N_{ev}/(\varepsilon \cdot N_{ann}\cdot B)$, where B is fraction of
$\phi$ decays into $K^+ K^-$. Total number of annihilations in the target
$N_{ann}$ is equal to $33.6\cdot10^6$ for gas and $10.6\cdot10^6$ for
liquid for the samples obtained without "slow particle" condition and 
$130.8\cdot10^6$ for gas and
$494.3\cdot10^6$ for liquid for the samples obtained with "slow particle" 
trigger. 
The   
yield values are affected by a
systematic uncertainty of about 6\%. It is due to
uncertainties of 
different selection procedures, particle identification, evaluation of the
apparatus efficiency and the total number of annihilated antiprotons.
In addition, the total yields are underestimated due to low acceptance
of $\phi\rho$ channel.

 The obtained total yield of the  $\omega\pi^+\pi^-$ channel:
$Y=(719\pm74)\cdot 10^{-4}$ for liquid and
$Y=(628\pm34)\cdot 10^{-4}$ for gas
are in general agreement with the results of previous measurements:
$Y=(660\pm60)\cdot 10^{-4}$ for liquid \cite{Biz.69} and
$Y=(682\pm74)\cdot 10^{-4}$ for gas \cite{Wei.93}.
The yields of $\phi\pi^+\pi^-$ channel:
$Y=(3.5\pm0.4)\cdot 10^{-4}$ for liquid and
$Y=(3.7\pm0.5)\cdot 10^{-4}$ for gas
are lower than those measured by other authors:
$Y = (4.6\pm0.9)\cdot 10^{-4}$ for liquid \cite{Biz.69} and
$Y = (5.4\pm1.0)\cdot 10^{-4}$ for gas \cite{Rei.91}. The reason is 
the lack of the events from $\phi\rho$ channel discussed above.
It is important to stress that this lack of $\phi\rho$ events is
not relevant to our subsequent analysis which based mainly on the
measurements of $\phi\pi\pi$ yield at small dipion masses.

\vspace{0.5cm}
As one can see from the Table,  
for all events, without any cut on dipion invariant mass,
the ratio $R(\phi/\omega)$ is about $5\cdot10^{-3}$
both in the case of annihilation in
liquid and in gas.
It is 
in agreement with the OZI rule prediction 
$R(\phi/\omega) \approx 4\cdot10^{-3}$. 
However, for
small masses of the dipion system 
the values of $R(\phi/\omega)$ increase by 3-4 times
leading to a significant 
deviation from the OZI rule prediction.

	This effect was observed for the first time 
and it is important to demonstrate
that it is not an artefact due to a difference 
in phase space
for $\phi$ and $\omega$ production.
The last column of Table  contains the ratios R corrected
for the difference in the phase space V for the 
reactions of $\phi\pi\pi$ and $\omega\pi\pi$ production.
$$R_C (\phi/\omega) = R(\phi/\omega)\frac{V(\omega\pi\pi)}{V(\phi\pi\pi)}$$
It turns out that after this correction the value of R only increases.
It should be stressed that  
different forms of such correction have been
discussed (see \cite{Rei.91},\cite{Ams.92} and references therein),
including those with corresponding scaling factor about 1. 
So one could not conclude which ratio $R$ or $R_c$ is more correct,
we will use whenever the $R_c$ values.

The increasing of R at small dipion masses for annihilation in liquid
could be explained as a manifestation of the difference
in the amplitudes of $\phi$ and $\omega$ production.
The conservation of C and P-parities in strong interactions
unambigously couples the spin-triplet $^3S_1$
initial state with the $\phi(\omega)\pi\pi$ final state when two pions are
in S-wave relative to each other. The spin-singlet initial $^1S_0$ state
couples with the $\phi(\omega)\pi\pi$ system when the two pions are in
P-wave, i.e. the  final state is $\phi(\omega)\rho$.
The partial wave
analysis of the $\phi\pi\pi$ channel 
demonstrates \cite{Tre.95} that in S-wave the
production of $\phi$ mesons is completely dominated by the spin-triplet
$^3S_1$ initial state,
whereas for the $\omega$ meson production both the $^3S_1$ and $^1S_0$ states
are important \cite{Biz.69}. So the ratio R for annihilation
in liquid, neglecting by the annihilation from the P-waves, is
\begin{equation}
R = \frac{Y(\phi\pi^+\pi^-)} {Y(\omega\pi^+\pi^-)} \approx
    \frac{Y(\phi(\pi^+\pi^-)_S)} {Y(\omega(\pi^+\pi^-)_S)+Y(\omega \rho) }
\end{equation}
It is clear that at small dipion masses, far from the $\rho$ peak,
this ratio should increase.

	Analogous consideration 
for annihilation in
gas is more complicated due to the significant contribution 
of the annihilation from 
P-wave. However, it is important to stress that  
the ratio R at small dipion masses in gas is high not 
due to 
increasing of the $\phi \pi^+\pi^-$  yield. 
As one could see from Table ,
the $Y(\phi \pi^+\pi^-)$ is the same for annihilation in liquid and in gas
at small dipion mass,
but it is the $\omega \pi^+\pi^-$ yield for $300 <M_{\pi\pi}<500~ MeV/c^2$
mass interval in gas to be less than in liquid. 
Therefore again, as in case of annihilation in liquid, we observed the
difference in the $\phi$ and $\omega$ production dynamics.

 In Fig.3 the  values of the $R_c$ 
are compared with the results
of other measurements of binary reactions of
antiproton annihilation at rest.
Some apparent dependence of
the degree of the OZI-rule violation on the mass  of the system created
with $\phi$ is seen.
For annihilation at rest  $\bar{N}N \to \phi X$ a decreasing of the
mass of X means an increase the momentum transferred to the $\phi$. 
However, one should be cautious, because different binary reactions 
were measured for different
initial states. 
Thus, the $\phi\gamma$,$\phi \rho$ and $\phi \omega$
 final states for annihilation from S-wave 
produced from the $^1S_0$ initial state.
Whereas the $\phi\pi$ and   
$\phi \eta$
channels come from the $^3S_1$ initial state.  As discussed above,
in liquid at small dipion masses both 
 $\phi \pi^+\pi^-$  and $\omega \pi^+ \pi^-$ channels 
come dominantly from the $^3S_1$ state.
Therefore the corresponding ratio R could be directly 
compared with  the values  of  $R(\phi\pi/\omega\pi)$ and
$R(\phi \eta /\omega \eta)$. Doing so,
it turns out that for the $\phi$ production from the $^3S_1$ 
initial state,  
the degree of the OZI rule violation 
increases with
decreasing of the mass of the system created with $\phi$, i.e. with
the increasing of the momentum transfer.

It would be interesting to investigate further this possible dependence 
of the degree of the OZI rule violation
on the
momentum transfer.
The same effect was found in $\phi$ production in 
$\pi^{\pm} N \to \phi N$ interaction
\cite{Coh.77} where the $d\sigma/dt$ distribution 
of $\phi$  production at large $t$ differs significantly from the 
one for $\omega$-meson, leading to the increase of $\phi/\omega$ ratio
at large $t$.
The direct measurements
of the t-dependence of the differential cross sections of $\phi\pi$ and
$\omega\pi$ channels  in $\bar{p}p$ annihilation in flight should
clarify the problem.

	In conclusion, the
measurement of the ratio $R = Y(\phi\pi^+\pi^-) / Y(\omega\pi^+\pi^-)$ for
annihilation of stopped antiprotons in gaseous and liquid hydrogen targets
was performed. It was found that the value of this ratio 
increases with the decreasing
of the dipion mass, which demonstrates the difference in the $\phi$ and
$\omega$ production mechanisms. 
An indication on the momentum transfer dependence of the apparent
OZI rule violation for  
$\phi$ production from the $^3S_1$ 
initial state was found.

\vspace{0.5cm}
	We would like to thank the technical staff of the LEAR machine
group for their support during the runs.

	We are very grateful to J.Ellis and D.Kharzeev for interesting
discussions.

	The JINR group acknowledges the support from  the
International Science Foundation, grant No. ML9300.

\newpage
Table . Yields of $\phi$ and $\omega$ meson production in reactions
(1)-(2) for annihilation of stopped antiprotons in liquid and gas targets
(only statistical errors are taken into account). $ R(\phi/\omega)$ and
$ R_C(\phi/\omega)$ are, respectively, the ratio of the yields without
and with phase space correction.

\begin{center}
\begin{tabular}{|c|cc|c|c|c|} \hline
& \multicolumn{2}{|c|}{ } & & & \\
& $ N_{ev}$ & $ \varepsilon,\%$ &
Y$\cdot 10^{4}$ & $ R(\phi/\omega)\cdot 10^3$ &
$ R_C (\phi/\omega)\cdot 10^3$ \\ \hline \hline
\multicolumn{6}{|c|}{Liquid H$_2$} \\ \hline \hline
\multicolumn{6}{|c|}{All events} \\ \hline
${\phi\pi^+\pi^-}$ & 853$\pm$94 & $1.08\pm0.03$ & 3.5$\pm$0.4 &
$ 4.9\pm 0.8$ & $ 10.3\pm 1.6$\\ \cline{1-4}
${\omega\pi^+\pi^-}$ & 32042$\pm$1541 & 4.60$\pm$ 0.41 & 719$\pm$74 & & \\
\hline
\multicolumn{6}{|c|}{$300~ MeV/c^2 < M_{\pi\pi} < 500~ MeV/c^2$} \\ \hline
${\phi\pi^+\pi^-}$ & 408$\pm$67 & $2.29\pm0.09$ & 0.7$\pm$0.1 &
$ 16.5\pm 3.5$ & $ 20.4\pm 4.3$ \\ \cline{1-4}
$\omega\pi^+\pi^-$ & 1918$\pm$244 & 4.67$\pm$ 0.40 & 42.4$\pm$6.5 & & \\
\hline \hline
\multicolumn{6}{|c|}{H$_2$ at 3 atm} \\ \hline \hline
\multicolumn{6}{|c|}{All events} \\ \hline
${ \phi\pi^+\pi^-}$ & 292$\pm$43 & 1.24$\pm$0.02 & 3.7$\pm$0.5 &
$ 5.9\pm 0.9$ & $ 12.5\pm 2.0$ \\ \cline{1-4}
${ \omega\pi^+\pi^-}$ & 71640$\pm$1011 & 3.80$\pm$ 0.20 & 628$\pm$34 & & \\
\hline
\multicolumn{6}{|c|}{$300~ MeV/c^2 < M_{\pi\pi} < 500~ MeV/c^2$} \\ \hline
${ \phi\pi^+\pi^-}$ & 104$\pm$23 & 2.41$\pm$0.06 & 0.7$\pm$0.2 & $29.3\pm 8.6$
&$ 36.2\pm 10.6$ \\ \cline{1-4}
${ \omega\pi^+\pi^-}$ & 2701$\pm$578 & 3.74$\pm$ 0.26 & 23.9$\pm$5.4 & & \\
\hline
\end{tabular}
\end{center}
\newpage
\begin{figure}
\psfull
\epsfig
{file=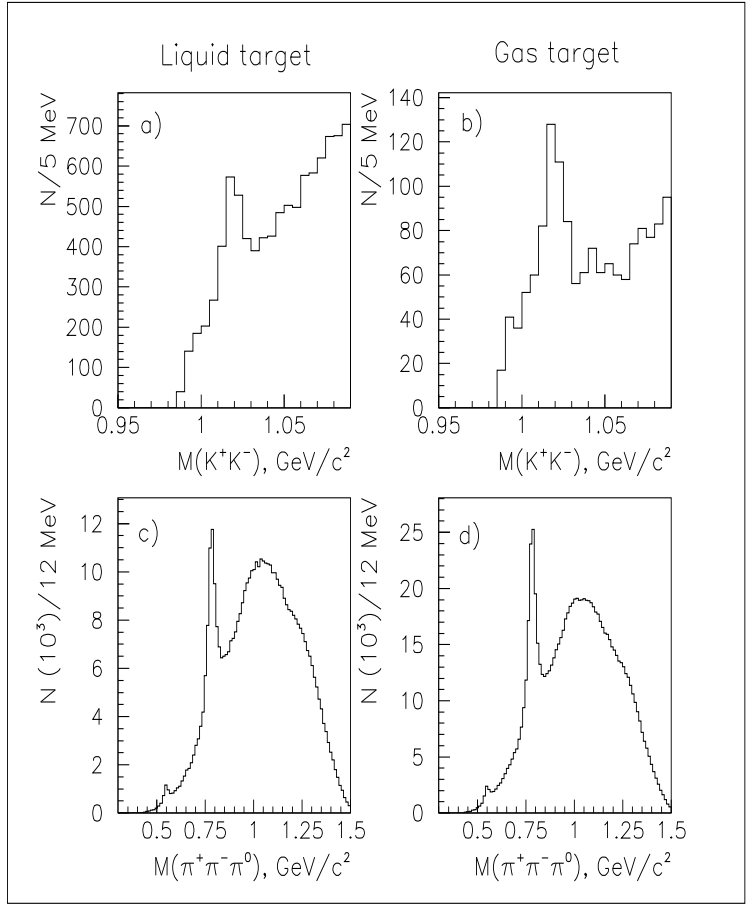,width=9cm,height=11cm,bbllx=-5cm,bblly=5cm,bburx=13.cm,bbury=23cm}
\caption{The $K^+K^-$ and $\pi^+\pi^-\pi^0$ invariant mass spectra for the
reactions $\bar pp\to K^+K^-\pi^+\pi^-$ (a,b) and 
$\bar pp\to 2\pi^+2\pi^-\pi^0$ (c,d) fr antiproton
annihilation at rest in liquid target (a,c) and gas target (b,d)
(uncorrected for apparatus acceptance) .}
\end{figure}

\begin{figure}
\psfull
\epsfig
{file=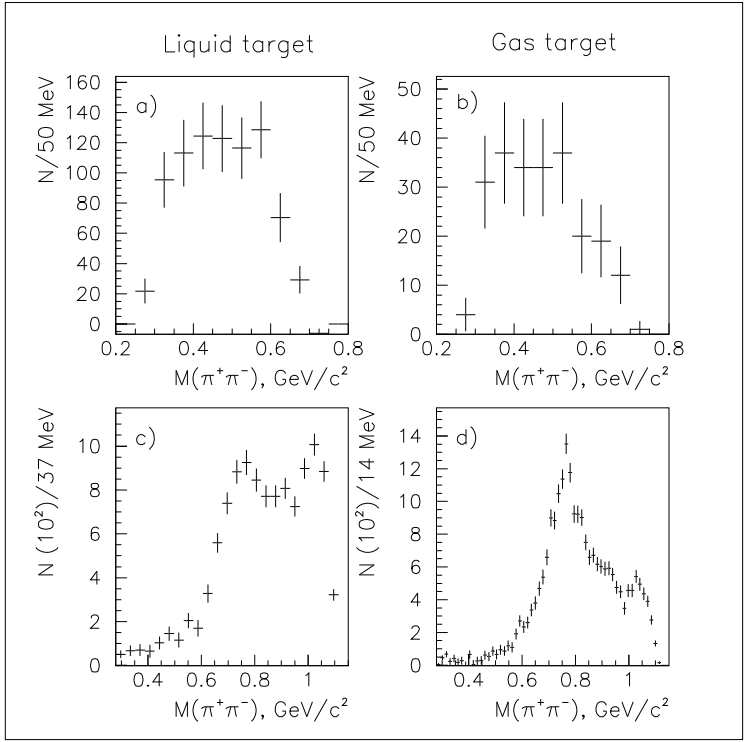,width=9cm,height=9cm,bbllx=-3cm,bblly=5cm,bburx=15cm,bbury=23cm}
\caption{Background free $\pi^+\pi^-$ invariant mass distributions for the
 reactions $\bar pp\to \phi\pi^+\pi^-$
(a,b) and $\bar pp\to \omega\pi^+\pi^-$ (c,d) for 
antiproton annihilation at rest in liquid target(a,c) and
gas target (b,d)
(uncorrected for apparatus acceptance) .}

\end{figure}

\begin{figure}[h]
\psfull
\epsfig
{file=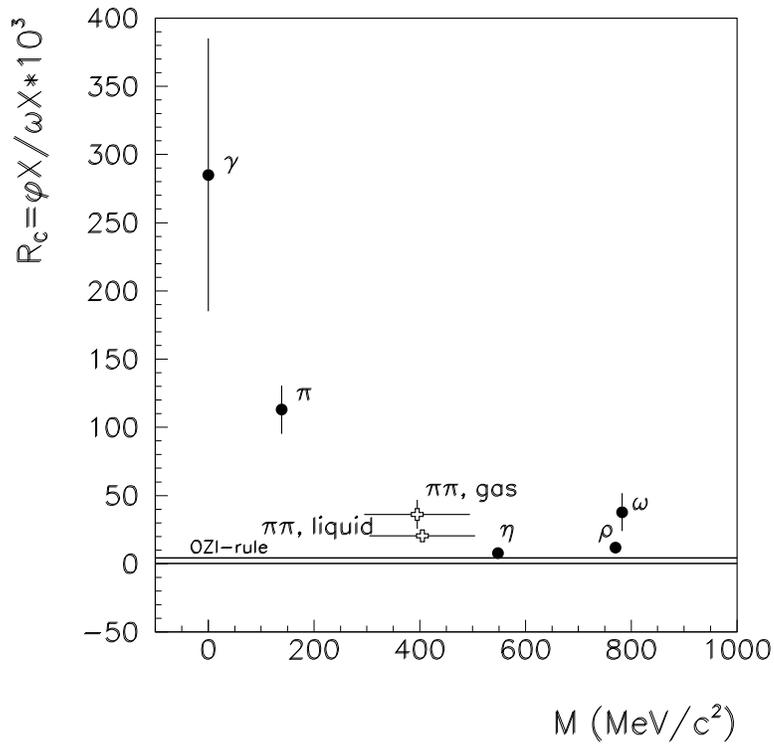,width=10cm,height=10cm,bbllx=-3cm,bblly=5cm,bburx=15cm,bbury=23cm}
\caption{The ratio $R_C=\phi X/\omega X $ for different reactions of $\bar pp$
 annihilation at rest as
a function of the mass M of the system X. The crosses indicate the result of 
this experiment.}
\end{figure}

\end{document}